\documentclass[aps,prl,twocolumn,groupedaddress]{revtex4}
\usepackage{graphicx}
\begin{document}
\title{Antiferromagnetic Spin Fluctuation above the Superconducting Dome and the Full-Gaps Superconducting State in LaFeAsO$_{1-x}$F$_x$ Revealed by $^{75}$As-Nuclear Quadrupole Resonance}
\author{T. Oka$^1$}
\author{Z. Li$^2$}
\author{S. Kawasaki$^1$}
\author{G. F. Chen$^2$}
\author{N. L. Wang$^2$}
\author{Guo-qing Zheng$^{1,2}$}
\affiliation{$^1$Department of Physics, Okayama University, Okayama 700-8530, Japan} 
\affiliation{$^2$Beijing National Laboratory
for Condensed Matter Physics, Institute of Physics, Chinese Academy of Sciences, Beijing 100190, China}
\date{\today}

\begin{abstract}
We report a systematic study by $^{75}$As nuclear-quadrupole resonance in LaFeAsO$_{1-x}$F$_{x}$.  The antiferromagnetic spin fluctuation (AFSF) found above the magnetic ordering temperature  $T_N$ = 58 K   for $x$ = 0.03 persists in the regime 0.04 $\leq x \leq$ 0.08 where superconductivity sets in.  A dome-shaped $x$-dependence of the superconducting transition temperature $T_c$ is found, with the highest $T_c$ = 27 K  at $x$ = 0.06 which is realized under  significant AFSF. With  increasing $x$ further, the AFSF decreases, and so does $T_c$. These features resemble closely the cuprates La$_{2-x}$Sr$_x$CuO$_4$. 
 In $x$ = 0.06, the spin-lattice relaxation rate ($1/T_1$) below $T_c$ decreases exponentially down to 0.13 $T_c$, which unambiguously indicates that the energy gaps are fully-opened.  The temperature variation of $1/T_1$ below $T_c$ is rendered nonexponential for other $x$ by impurity scattering.

\end{abstract}

\maketitle
The discovery of superconductivity in LaFeAsO$_{1-x}$F$_x$ at the transition temperature $T_c$ = 26 K \cite{Kamihara} has gained much attention in the condensed-matter physics community. The electron -doping (F -doping) suppresses the antiferromagnetic ordering at $T_N$ = 140 K in LaFeAsO and high-$T_c$ superconductivity appears\cite{Kamihara}. The $T_c$ significantly increases up to 55 K in RFeAsO$_{1-x}$F$_x$ ($R$: Ce, Pr, Nd, Sm) \cite{XHChen,ZARen}. 
 To elucidate the mechanism of Cooper pairs formation in these arsenides, it is essential to know the superconducting gap symmetry and the normal-state properties. Previous nuclear-magnetic resonance (NMR) and nuclear-quadrupole resonance (NQR) measurements have found that the superconductivity is in the spin-singlet state with multiple gaps \cite{MatanoPr,KawasakiLa,MatanoBa122}. Recent systematic measurements on Ba(Fe$_{1-x}$Co$_x$)$_2$As$_2$\cite{Ahilan}, CaFe$_2$As$_2$ under pressure\cite{KawasakiCa}, LaNiAsO$_{1-x}$F$_{x}$\cite{Tabuchi}, and BaFe$_2$(As$_{1-x}$P$_x$)$_2$\cite{Nakai_P122} have suggested that the antiferromagnetic spin fluctuation (AFSF) originated from their multiple electronic bands correlates with the appearance of  the pertinent superconducting properties.  
On the other hand, there are also reports suggesting that AFSF is not important to realize high $T_c$.\cite{Kinouchi}

For prototypical LaFeAsO$_{1-x}$F$_x$, several issues remain elusive. One is the role of AFSF. In cuprates, it has been believed that AFSF plays a crucial role to induce high-$T_c$ superconductivity, but the situation in
  LaFeAsO$_{1-x}$F$_x$ is still unclear \cite{KawasakiLa,Nakai,Nakai2,Mukuda,Grafe,Kobayashi2}. 
Some previous studies by NMR found no AFSF \cite{Nakai,Nakai2,Mukuda,Grafe}.

The second issue is the doping dependence of $T_c$. It was  initially reported that $T_c$ forms a wide plateau at 0.04 $\leq x \leq$ 0.12 \cite{Kamihara}, which raises a question about the effect of doping.
The third unresolved issue is the  superconducting gap symmetry.
The spin-lattice relaxation rate (1/$T_1$) decreases sharply below $T_c$, but the data were insufficient for distinguishing between $d$-wave from sign-reversal $s$-wave \cite{KawasakiLa,Nakai,Nakai2,Mukuda,Grafe,Kobayashi2}. From other experimental probes, some measurements suggested the existence of node \cite{Martin}, but  the photoemission spectroscopy and the point contact Andreev reflection measurement suggested a nodeless gap \cite{Sato,Gonnelli}.

Here we report results of systematic $^{75}$As NQR studies on LaFeAsO$_{1-x}$F$_{x}$  ($x$ = 0.03, 0.04, 0.06, 0.08, 0.10, and 0.15). 
An  antiferromagnetic order with $T_N$ = 58 K is found for $x$ = 0.03. Bulk superconductivity sets in at $T_c$ = 21 K for $x$ = 0.04, with strong  AFSF. 
 A dome-shaped $x$-dependence of $T_c$ is found, with the highest $T_c$ = 27 K  at $x$ = 0.06 which is realized under significant AFSF.
With further doping, the AFSF is weakened   and disappears for $x \geq$ 0.10.
Concomitantly, $T_c$ decreases. These features resemble closely the case of cuprates La$_{2-x}$Sr$_x$CuO$_4$, and suggests that the AFSF is important in producing the superconductivity in LaFeAsO$_{1-x}$F$_{x}$ as well.
The systematic observation of the AFSF in the low-doping regime is unprecedented, and the high quality samples enable us to reveal a dome shape of the $T_c$ which has a maximum at quite low $x$.
In the superconducting state, $1/T_1$ for $x$ = 0.06 decreases $exponentially$ down to 0.13 $T_c$, which is clear and direct evidence for a fully -gapped superconducting state.
The $T$-variation of $1/T_1$ below $T_c$ is rendered nonexponential for  $x$ either smaller or larger than 0.06, showing a seemingly $T^3$ behavior for $x$ = 0.10, which is accounted for by  impurity scattering.

The polycrystalline samples of LaFeAsO$_{1-x}$F$_{x}$ ($x$ = 0.03, 0.04, 0.06, 0.08, 0.10, and 0.15) were synthesized by the solid state reaction method\cite{GFChen,SUP}. Here, $x$ indicates the nominal composition of the starting material.  
Quite often, resistivity measurements give a higher $T_c$ than magnetic susceptibility or NQR. We define $T_c$ by the latter methods.
ac susceptibility measurements using the in-situ NQR coil indicate  $T_c$ = 21, 27, 23, 18, and 12 K for $x$ = 0.04, 0.06, 0.08, 0.10, and 0.15, respectively. The $1/T_1$ decreases exactly below such-determined $T_c$.
The $T_1$ is determined by an excellent fit to the single exponential curve $1-\frac{M(t)}{M_0}$ = $\exp(\frac{-3t}{T_1})$\cite{SUP}, where $M_0$ and $M(t)$ are the nuclear magnetization in the thermal equilibrium and at a time $t$ after the saturating pulse, respectively.

\begin{figure}[h]
\includegraphics[width=8.5cm]{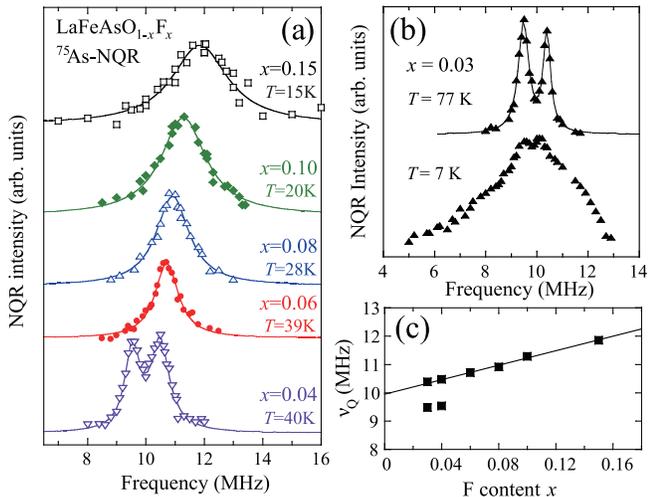}
\caption{\label{fig1} (color online)
(a) Doping dependence of the $^{75}$As-NQR spectrum for LaFeAsO$_{1-x}$F$_{x}$ measured above $T_c$.  Data for $x$ = 0.08 are from Ref.\cite{KawasakiLa}. Solid curves are Lorentzian fittings which give a FWHM of $\sim$ 0.95, 1.2, 1.8, and 2.3 MHz for $x$ = 0.06, 0.08, 0.10, and 0.15, respectively. (b) The spectra above and below $T_N$ = 58 K for $x$ = 0.03. (c) The $x$ dependence of $\nu_Q$.}
\end{figure}

Figure 1 (a) shows the $^{75}$As-NQR spectrum for 0.04 $\leq x \leq$ 0.15 measured above $T_c$. 
As seen in the figure, a clear single peak, which can be fitted by a single Lorentzian curve, is observed for $x \geq$ 0.06. The spectra do not change below $T_c$. 
However, we observed two peaks for $x$ = 0.03 (Fig. 1(b)) and 0.04. This indicates that there are two As -sites which are in different surroundings.
The NQR frequency $\nu _Q$ increases with increasing $x$, as seen in Fig. 1(c). Here, $\nu_{\rm Q}$ probes the electric-field gradient generated by the carrier distribution and the lattice contribution surrounding the As nucleus. The doping evolution of $\nu_Q$, the spectral shapes, and the single component of $T_1$ indicate that the electron carriers were homogeneously doped for $x \geq$ 0.06, but phase separation occurs in $x$ = 0.03  and 0.04. We speculate that this may be due to the local distribution of the F ion around the As nucleus, which is inevitable in a quite  low-doping region. 
 Remarkably, the $T$ dependences of $1/T_1$ measured at each peak of $x$ = 0.03  and 0.04 indicate that each phase is homogeneous.  
The same behavior was found in Ref.\cite{Grafe2}, where the $\nu_Q$ is quite similar to ours although the nominal $x$ there is  larger than ours by $\sim$ 0.02\cite{SUP}. 
Figure 1(b) shows the $T$ evolution of the NQR spectra for $x$ = 0.03. Below $T_N$ = 58 K, the spectra are broadened due to an antiferromagnetic order.

Figures 2(a) and 2(b) show the $T$ dependence of $1/T_1$ for all samples.  
For  $x$ = 0.03,   $1/T_1$ shows a small upturn right above $T_N$ and then decreases below, leaving a tiny peak at $T_N$. 
For $x \geq$ 0.04, $1/T_1$ decreases rapidly below $T_c$ due to the opening of the superconducting energy gaps.

 \begin{figure}[h]
\includegraphics[width=8cm]{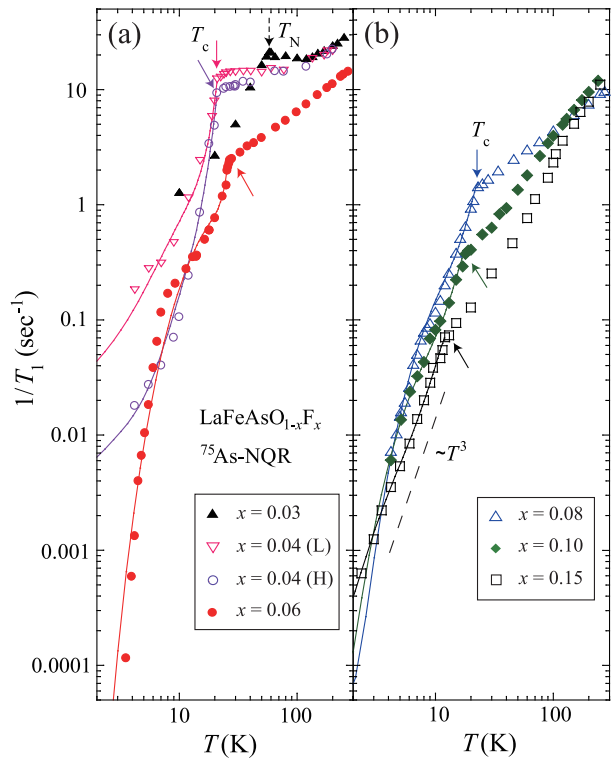}
\caption{\label{fig1} (color online) The $T$ dependences of $1/T_1$ for $x$ = 0.03, 0.04 and 0.06 (a) and for $x$ = 0.08, 0.10, and 0.15 (b). Data for $x$ = 0.03 were collected at the high-frequency (H) NQR peaks. For $x$ = 0.04,  $1/T_1$ was measured at both the low-frequency (L) and the H peak.  
Solid curves below $T_c$ for $x \geq$ 0.04 are the simulations based on a $s^{\pm}$ wave superconducting gap model with impurity scattering (see the text). The dashed line indicates the relation $1/T_1$ $\propto$ $T^3$. The dotted and solid arrows indicate $T_N$ and $T_c$, respectively.}
\end{figure}

\begin{figure}[h]
\includegraphics[width=8.5cm]{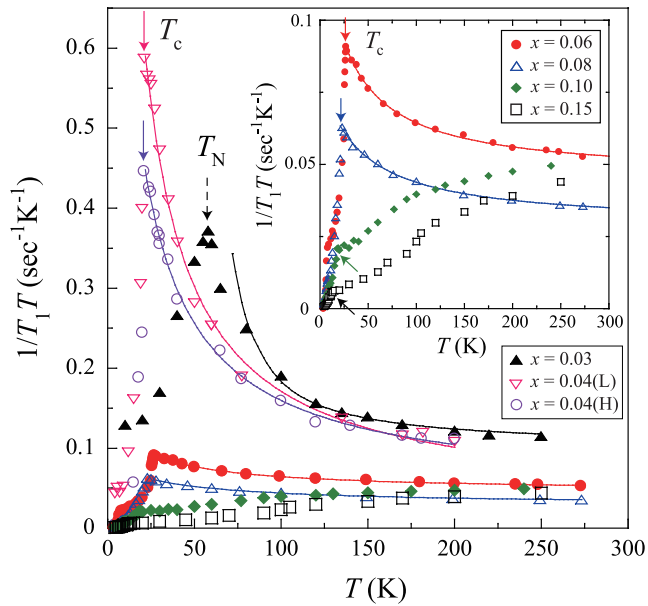}
\caption{\label{fig1} (color online) $T$ dependence of 1/$T_1T$ for various $x$. The curves above $T_N$ or $T_c$ are fits to the AFSF theory (see the text). 
The inset is the enlarged part for 0.06 $\leq x \leq$ 0.15.}
\end{figure}

Before going into the detail of the superconducting state, we first discuss the normal-state property. For this purpose, we plot $1/T_1T$  {\it vs} $T$ in Fig. 3. None of the samples shows a Korringa relation 1/$T_1T$ = const. expected for a conventional metal. 
Above $T_N$ of  $x$ = 0.03, $1/T_1T$ increases with decreasing $T$ due to the AFSF. 
Such AFSF persists in  $x$ = 0.04, 0.06, and 0.08, where $1/T_1T$ increases with decreasing $T$ down to $T_c$.  
To model the $1/T_1T$ above $T_N$ or $T_c$, we employed the theory for a weakly antiferromagnetically-correlated metal \cite{Moriya}, 1/$T_1T$ =(1/$T_1T)_{AF}+$(1/$T_1T)_0$= $C/(T+\theta)$ + (1/$T_1T)_0$. Here, the first term described the contribution from the antiferromagnetic wave vector, and the second term is the contribution from the density of states (DOS) at the Fermi level. For $x$ = 0.03, $\theta$ is simply $-T_N$,  where  the  data  can be well fitted except around $T_N$ \cite{Note}.  
As seen in Fig. 3,  1/$T_1T$ for $x$ = 0.04, 0.06, and 0.08  are well reproduced by this model with $\theta$ $\sim$ 10, 25 and 39 K, respectively. The low-frequency NQR peak  for $x$ = 0.04 gives a smaller $\theta$ $\sim$ 5 K.
The increase of $\theta$ with increasing $x$ means that the system moves away from the magnetic instability (MI) where $\theta$ = 0 K.
With further doping, for $x$ = 0.10 and 0.15, no enhancement of $1/T_1T$  is seen. Instead,  $1/T_1T$ decreases with decreasing $T$, which  was recently explained by the loss of the DOS due to a topological change of the Fermi surface \cite{Tabuchi,Ikeda}. 
The results in previous reports of the lack of the AFSF for $x \geq$ 0.10 \cite{Nakai,Nakai2,Grafe} are consistent with our results for $x$ = 0.10 and 0.15. 

The remarkable finding is that the highest $T_c$ = 27 K is realized at  $x$ = 0.06, which is  away from the MI. 
This situation is quite similar to the cuprates La$_{2-x}$Sr$_{x}$CuO$_4$\cite{Ohsugi}.
In the scenario of spin fluctuation-mediated superconductivity, this can be understood as follows. At high doping levels, the decrease of $T_c$ is due to the weakening of the AFSF. In the vicinity of the MI, on the other hand, the too strong low-energy fluctuation acts as pair breaking\cite{UedaMoriya}.   Therefore, a maximal $T_c$ is realized at some point away from the MI with moderate AFSF.

Figure 4 shows the phase diagram for LaFeAsO$_{1-x}$F$_{x}$ obtained in the present study. The most important finding is that the highest $T_c$ is found in the low-doping regime, which makes our $T_c$ vs $x$ relation look like a dome shape. 
In the previous study \cite{Kamihara,Luetkens}, the failure of obtaining higher $T_c$ in the low-doping regime is probably due to  sample inhomogeneity as evidenced by the broader (in fact, two-peak-featured)  NQR spectrum \cite{Nakai3}.  The present phase diagram is consistent with that for Ba(Fe$_{1-x}$Co$_x$)$_2$As$_2$\cite{Ahilan} but is somewhat different from that for BaFe$_2$(As$_{1-x}$P$_x$)$_2$\cite{Nakai_P122}, whose $T_c$ shows a maximum  around $\theta$ = 0.  This slight difference may originate from the difference of the tuning parameter for their ground states. The ground states for both LaFeAsO$_{1-x}$F$_{x}$ and  Ba(Fe$_{1-x}$Co$_x$)$_2$As$_2$ are tuned by electron doping. On the other hand, isovalent P doping acts as chemical pressure on BaFe$_2$(As$_{1-x}$P$_x$)$_2$.  In any case, these phase diagrams support the intimate relationship between AFSF and superconductivity in iron arsenides.  Furthermore, such a phase diagram has consistently been found in high-$T_c$ cuprate La$_{2-x}$Sr$_{x}$CuO$_4$\cite{Ohsugi} and heavy fermion compounds\cite{Lonzarich}, indicating that the AFSF plays a significant role to induce superconductivity in strongly correlated electron systems in general.

\begin{figure}[h]
\includegraphics[width=7.5cm]{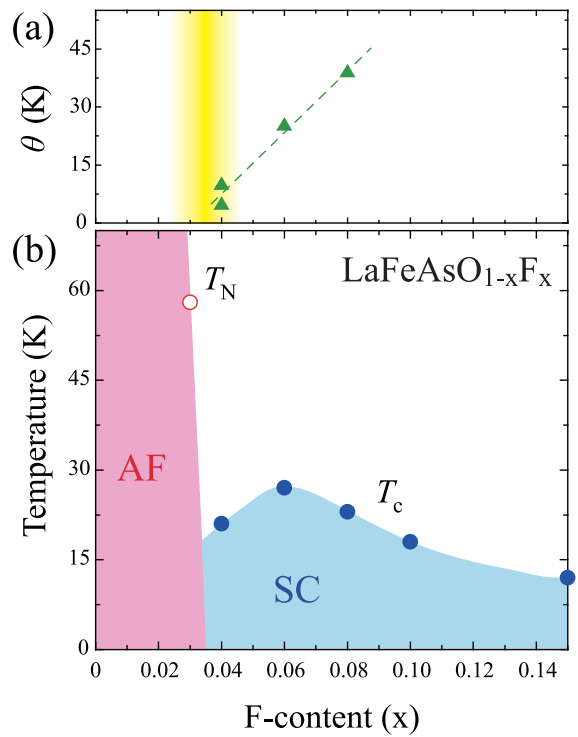}
\caption{\label{fig1} (color online) Phase diagram obtained in this study. AF and SC denote the antiferromagnetically ordered and superconducting states, respectively. (a) $x$ dependence of $\theta$. The dotted line is a guide to the eyes. The shade indicates the region of phase separation. (b) $x$ dependence of $T_N$ and $T_c$ determined by NQR measurements.}

\end{figure}


 Next, we turn to the superconducting state. Figure 5(a) shows the $T$ dependence of $1/T_1$ for $x$ = 0.06.
Below $T_c$, $1/T_1$ decreases steeply due to the opening of the superconducting gaps. The hump structure at $T \sim$ 0.4 $T_c$ is due to the multiple-gap character as reported for other compounds\cite{MatanoPr,KawasakiLa,MatanoBa122,KawasakiCa}. The $T$ variation at low $T$ is much stronger than $T^3$, and even stronger than $T^5$, as can be clearly seen in the figure. In fact, $1/T_1$ decreases exponentially below 0.4 $T_c$. In Fig. 5(b), we plotted $1/T_1$ against $T_c/T$ in a semilogarithmic scale. As indicated by the solid line, the $1/T_1$ below $T \sim$ 0.4 $T_c$ clearly follows the relation $1/T_1$ $\propto$ exp($-\Delta_0$/$k_BT$) with $\Delta_0$/$k_BT_c$ = 1.8, where $\Delta_0$ and $k_B$ denote the gap size at $T$ = 0 and the Boltzmann constant, respectively.  
 This is clear and direct evidence that the superconducting state is fully gapped in LaFeAsO$_{0.94}$F$_{0.06}$.  
 
 \begin{figure}[h]
\includegraphics[width=8.5cm]{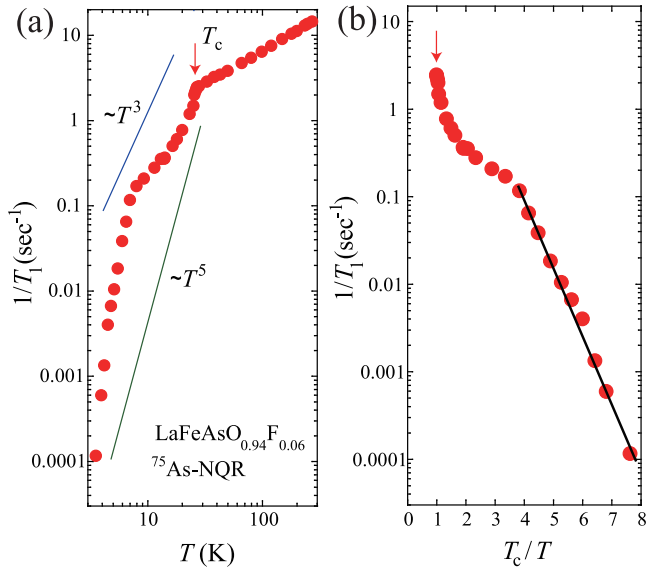}
\caption{\label{fig1} (color online) (a) The $T$ dependence of $1/T_1$ for $x$ = 0.06.  (b) Semilogarithmic plot of $1/T_1$ vs $T_c/T$.  The solid line represents the relation $1/T_1$ $\propto$ exp($-\frac{\Delta_0}{k_BT}$). }

\end{figure}

The evolution of the superconducting-state properties can be seen in Fig. 2. For $x$ = 0.06-0.10, $1/T_1$ shows a marked hump structure around $T \sim$ 0.4 $T_c$ and is followed by a still sharper decrease below. However, the low-$T$ behavior of $1/T_1$ changes gradually, as to decrease less and less steeply as $x$ increases. Eventually, for $x$ = 0.15, the hump structure disappears completely. Instead, a simple $T$ dependence emerges which is close to $T^3$. Such $T^3$ behavior has been reported previously \cite{Nakai,Mukuda,Grafe} and was taken as evidence for line nodes. Below we show that it is a consequence of impurity scattering.
 Namely, the $T^3$ is an accidental one rather than an intrinsic one. In fact, in Ba$_{1-x}$K$_x$Fe$_2$As$_2$, the low-$T$ behavior of $1/T_1$ also changes when the sample purity differs \cite{MatanoBa122,LiBa122}. 

Assuming sign reversing $s$-wave symmetry \cite{Mazin,Kuroki} with impurity scattering, one can reproduce the evolution of the $1/T_1$ below $T_c$.
By introducing the impurity scattering parameter $\eta$ in the energy spectrum in the form of $E = \omega + i \eta$, the $1/T_1$ in the superconducting state is given by $\frac{T_1(T_c)}{T_1(T)} \cdot \frac{T_{\rm c}}{T} = \frac{1}{4T} \int _{-\infty}^{\infty}
\frac{{\rm d} \omega}{{\rm cosh}^{2} \frac{\omega}{2T}}(W_{\rm GG}+W_{\rm FF})$  \cite{LiLiFeAs}  where $W_{\rm GG} = [\langle{\rm Re}\{(\omega +i \eta)/\sqrt{(\omega +i \eta)^{2}+|\Delta(k_{\rm F})|^{2}} \}\rangle_{k_{\rm F}}]^{2}$  and  $W_{\rm FF} = [\langle{\rm Re}\{ 1/\sqrt{(\omega +i \eta)^{2}+|\Delta (k_{\rm F})|^{2}}\} \Delta (k_{\rm F})\rangle_{k_{\rm F}}]^{2}$. Here the $\Delta$ is the gap parameter, and $\left< \dots \right>$ is the average over the entire Fermi surface, and runs over three bands consisting of two hole pockets at the $\Gamma$ point and an electron pocket at the $M$ point, respectively \cite{DJSingh}. Namely, for a quantity $F$, $\left< F\left[\Delta (\bf k_{\rm F})\right]  \right> _{\bf k_{\rm F}} = \left[N_{1} F(\Delta^+_{1}) +N_{2} F(\Delta^-_2)+N_{3} F(\Delta^-_{3})\right]/(N_{1} + N_{2}+ N_{3}$), where $N_{i}$ is the DOS coming from band $i$ ($i = 1, 2, 3$). 
Here, it is tempting to assign bands $1$, $2$, and $3$ to the $\gamma$, $\beta$, and $\alpha$ bands found in angle-resolved photoemission spectroscopy measurement \cite{Ding}. 
It is noted that the weaker $T$ dependence in the $x$ = 0.15 sample can be understood as due to the impurity scattering that brings about a finite DOS.  For $x$ = 0.04 where two As sites were found, $1/T_1$ for each site can also be fitted by the same model, with an additional feature that a large $\eta$ is needed to explain the low-$T$ behavior. This can be understood if the two phases coexist in the nanoscale \cite{Grafe2}, where one phase acts as an impurity scatterer for the other. The obtained fitting parameters are summarized in Table 1.
 Finally, we note that an $s^{++}$ wave \cite{Kontani} seems difficult to explain the lack of the coherence peak just below $T_c$ and the $x$ evolution of low-$T$ behavior of $1/T_1$. 

\begin{table}
\centering
\caption[]{\footnotesize The fitting parameters $\Delta_1^+$ ( = $\Delta_3^-$), $\Delta_{2}^{-}$, $\eta$ in the unit of $k_BT_c$, and $N_{1}$:$N_{2}$:$N_{3}$.}

\begin{tabular}{c|ccccc}
\hline $x$ & $T_c$ (K) & $\Delta^+_1$  & $\Delta^-_2$ & $\eta$ & $N_1$:$N_2$:$N_3$  \\ 
\hline 0.04(L) &  21  & 4.50 & 0.93 & 0.39 & 0.335:0.330:0.335 \\
0.04(H) &  21  & 4.58 & 1.63 & 0.27 & 0.38:0.24:0.38 \\
0.06 &  27  & 5.62 & 1.11 & 0.006 & 0.30:0.40:0.30 \\ 
 0.08 &  23 & 3.37 & 0.92  & 0.03  &  0.303:0.394:0.303\\ 
  0.10 &  18 & 3.00 & 0.83 & 0.035  &  0.305:0.39:0.305 \\ 
 0.15 &  12  & 2.62 & 0.79 & 0.15 & 0.31:0.38:0.31 \\ 
\hline 
\end{tabular}

\label{table1}
\end{table}

In conclusion, we have presented the results of systematic NQR measurements on high quality samples of LaFeAsO$_{1-x}$F$_{x}$ ($x$ = 0.03, 0.04, 0.06, 0.08, 0.10, and 0.15). 
The AFSF seen above $T_N$ = 58 K of $x$ = 0.03 persists in the 0.04 $\leq x \leq$ 0.08 regime.
The highest $T_c$ = 27 K  is realized for $x$ = 0.06 which is  away from the magnetic instability but with significant AFSF. The phase diagram closely resembles  those of the cuprates La$_{2-x}$Sr$_x$CuO$_4$ and other iron arsenides, which suggests that the AFSF is also important to produce the superconductivity in LaFeAsO$_{1-x}$F$_{x}$. 
 In $x$ = 0.06, $1/T_1$ below $T_c$ decreases exponentially down to 0.13 $T_c$, which unambiguously indicates that the superconducting gaps are fully-opened.  The $T$-variation of $1/T_1$ below $T_c$ is rendered nonexponential for $x$ either smaller or larger than 0.06, which is accounted for by  impurity scattering.

We thank M. Ichioka for help in the calculation and K. Ishida for useful communication. 
Work in Okayama was supported in part by research grants from MEXT (No. 22103004 and 23102717). Work in Beijing was supported by CAS and NSFC. 

\begin{thebibliography}{}
\bibitem{Note}
The deviation may be understood as   due to impurity scattering by another phase that  coexists in the nanoscale \cite{Grafe2}.
\end{thebibliography}

\end{document}